\documentclass[twocolumn, pra, aps]{revtex4-2}

\usepackage{amsmath}
\usepackage{amsfonts}
\usepackage{bm}
\usepackage{graphicx}

\begin{document}

\title{Long lifetime supersolid in a two-component dipolar Bose-Einstein condensate}

\author{Shaoxiong Li}
\affiliation{Department of Engineering Science, University of
Electro-Communications, Tokyo 182-8585, Japan}

\author{Uyen Ngoc Le}
\affiliation{Department of Engineering Science, University of
Electro-Communications, Tokyo 182-8585, Japan}

\author{Hiroki Saito}
\affiliation{Department of Engineering Science, University of
Electro-Communications, Tokyo 182-8585, Japan}

\begin{abstract}
Recent studies on supersolidity in a single-component Bose-Einstein
condensate (BEC) have relied on the Lee-Huang-Yang (LHY) correction
for stabilization of self-bound droplets, which however involves a high
density inside the droplets, limiting the lifetime of the supersolid.
Here we propose a two-component mixture of dipolar and nondipolar
BECs, such as an $^{166}$Er-$^{87}$Rb mixture, to create and stabilize
a supersolid without the LHY correction, which can suppress the atomic
loss and may allow observation of the long-time dynamics of the
supersolid.
In such a system, supersolidity can be controlled by the difference in
the trap centers between the two components.
\end{abstract}

\date{\today}

\maketitle{}

A supersolid state is a peculiar state of matter that possesses both
the superfluidity of frictionless flow and the periodic spatial
structure of a crystal.
This counterintuitive state of matter was originally predicted to
arise in solid $^4$He~\cite{andreev1969quantum,leggett1970can}.
However, such a state of matter was first realized in Bose-Einstein
condensates (BECs) of ultracold atomic gases with a cavity-field
interaction~\cite{leonard2017}, spin-orbit interaction~\cite{li2017},
and magnetic dipole-dipole interaction
(DDI)~\cite{tanzi2019PRL,bottcher2019PRX, bottcher2019PRR,
  tanzi2019Nat, guo2019, natale2019, hertkorn2019, chomaz2019,
  norcia2021, tanzi2021, ilzhofer2021, sohmen2021, hertkorn2021PRL,
  hertkorn2021PRX, petter2021}.
In a BEC with a strong DDI, the matter wave splits into self-bound
droplets, which form crystal structures~\cite{kadau2016}.
Unlike incoherent isolated droplets simply placed in order, in a
supersolid state, the matter-wave droplets allow superflow between
them, giving rise to phase coherence.
Due to the presence of a crystal structure with coherence, intriguing
quantum properties emerge~\cite{baillie2018, zhang2019, roccuzzo2019,
  roccuzzo2020, gallemi2020, blakie2020, tengstrand2021,
  ancilotto2021, zhang2021, poli2021, hertkorn2021PRR, turmanov2021,
  pal2022, roccuzzo2022, roccuzzo2022prr}, such as a nonclassical
moment of inertia~\cite{tanzi2021} and low-energy Goldstone
modes~\cite{tanzi2019Nat, guo2019}.

In a self-bound droplet, the system is prone to collapse due to the
attractive part of the DDI.
The mean-field theory cannot explain the stability of the droplet
against collapse~\cite{xi2016, bisset2015}.
To explain the experimental observation of stable droplets the
beyond-mean-field effect, called the Lee-Huang-Yang (LHY)
correction~\cite{lee1957}, is needed.
It was found that the Gross-Pitaevskii (GP) equation with the LHY
correction can reproduce the experimental results
quantitatively~\cite{wachtler2016, wachtler2016_2, ferrier2016,
  saito2016}.
However, the LHY correction is a higher-order correction with respect
to the atomic density, and in order to arrest the collapse, the
density inside the droplet must be large.
As a result, atomic loss due to three-body recombination becomes
predominant, which limits the lifetime of the
droplets~\cite{schmitt2016,chomaz2016}.
Thus, the lifetime of the supersolid is restricted as long as the LHY
correction plays a dominant role in its stabilization.
Moreover, the use of the Feshbach resonance also increases the atomic
loss~\cite{chomaz2016}.
The limited lifetime of the supersolid hinders, for example, direct
observation of the slow oscillation of Goldstone modes~\cite{guo2019}.

To circumvent this problem, in the present paper, we propose to use a
two-component BEC.
In a two-component BEC, there exist two kinds of excitation modes:
density waves and quasi-spin waves.
When the inter- and intra-component interactions are comparable to
each other, the excitation energies of quasi-spin waves can be
smaller than those of density waves.
Consequently, the quasi-spin is more easily modulated than the total
density, and even a moderate strength of DDI can generate quasi-spin 
patterns without causing dipolar collapse.
Two-component mixtures of dipolar BECs have been experimentally
realized~\cite{trautmann2018, politi2022} and theoretically
investigated thoroughly~\cite{goral2002, saito2009, gligoric2010,
  jain2011, xi2011, young2012, wilson2012, young2013, zhao2013,
  adhikari2014, zhang2016, kumar2017, pastukhov2017, xi2018,
  kumar2019, lee2021}.
A self-bound droplet including two components was also
proposed~\cite{bisset2021, smith2021}.
However, there have been few studies on supersolidity in
two-component dipolar mixtures.
Very recently, supersolid formation in a two-component dipolar BEC was
theoretically studied~\cite{scheiermann2022}.
However, in this study, the periodic structure is formed in the total
density, not in the quasi-spin, and hence the LHY correction is
needed to stabilize the droplets~\cite{scheiermann2022}.

We use a $^{166}$Er-$^{87}$Rb mixture to study the two-component
supersolid state.
The magnetic moment of $^{166}$Er is much larger than that of
$^{87}$Rb, and the $^{166}$Er BEC forms droplets and a supersolid,
while the $^{87}$Rb BEC plays the role of a medium that accommodates
them.
The use of $^{166}$Er is suitable for the present purpose, because the
magnetic moment is not too large and the formation of LHY droplets
can be avoided.
Because of this moderate strength of DDI, for a single-component
$^{166}$Er BEC, reduction of the scattering length by Feshbach
resonance is needed to produce the droplets~\cite{chomaz2016}.
By contrast, in our two-component system, we will show that the
droplet lattice can be formed without reducing the scattering length,
which also suppresses the atomic loss and prolongs the lifetime.
We will show that one- and two-dimensional supersolid states can be
formed, and that they are stabilized without the LHY correction.
An important parameter in this system is the difference $\delta$ in
the trap centers between the two components, including gravitational
sag.
By controlling $\delta$, we can change the Josephson link between the
droplets, and then we can control the degree of supersolidity.
We will demonstrate that the long-period out-of-phase Goldstone mode
appears when the supersolidity is enhanced by $\delta$.

We employ the mean-field approximation at zero temperature to describe
BECs of two atomic species with masses $m_1$ and $m_2$ and
magnetic moments $\mu_1$ and $\mu_2$.
The magnetic moments of the atoms are fixed in the $z$ direction by
the magnetic field.
The dynamics of the macroscopic wave functions $\psi_j(\bm{r}, t)$
can be described by the coupled nonlocal GP equations
\footnote{Here we ignore the LHY term. We have numerically confirmed
that our results are almost unchanged even in the presence of the LHY
term.}
$(j = 1, 2)$,
\begin{eqnarray}
& & i \hbar \frac{\partial}{\partial t} \psi_j(\bm{r}, t) =
\biggl[ -\frac{\hbar^{2}}{2 m_j} \nabla^{2} + V_j(\bm{r})
    + \sum_{j'=1}^2 g_{jj'} |\psi_{j'}(\bm{r}, t)|^2 
  \nonumber \\ 
  & & + \frac{\mu_0 \mu_j}{4\pi} \int d\bm{r}'
  \frac{1 - 3 \cos^2\theta}{|\bm{r}-\bm{r}'|^3}
  \sum_{j'=1}^2 \mu_{j'} |\psi_{j'}(\bm{r}', t)|^2
  \biggr] \psi_j(\bm{r}, t),
  \label{GP}
\end{eqnarray}
where $g_{jj'} = 2\pi\hbar^2 a_{jj'} / m_{jj'}$ are the contact
interaction coefficients, $a_{jj'}$ and $m_{jj'}=(m_j^{-1} +
m_{j'}^{-1})^{-1}$ are the $s$-wave scattering lengths and the reduced
masses between components $j$ and $j'$, $\mu_0$ is the magnetic
permeability of the vacuum, and $\theta$ is the angle between $\bm{r}
- \bm{r}'$ and the $z$ direction.
The wave functions are normalized as $\int d\bm{r} |\psi_j|^2 = N_j$,
where $N_j$ is the number of atoms in component $j$.
We assume that the system is confined in an optical dipole trap with two
different laser frequencies, by which the trap frequencies for the two
components can be controlled individually.
When the centers of the harmonic potentials for the two components
horizontally coincide, the external potential can be written as
\begin{equation}
 V_j(\bm{r}) = \frac{1}{2} m_j \left[ \omega_{jx}^2 x^2 +
   \omega_{jy}^2 y^2 + \omega_{jz}^2 (z - \delta_j)^2 \right],
\end{equation}
where $\omega_{j x, y, z}$ are the trap frequencies for component $j$.
The deviations of the trap centers $\delta_j$ in the $z$ direction
arise both from the gravitational sag and the laser alignment, which
we simply call ``sag'' in the following.
We define
\begin{equation}
  \delta_1 - \delta_2 \equiv \delta.
\end{equation}

We assume that components 1 and 2 consist of $^{166}$Er atoms with
$\mu = 7 \mu_B$ and $^{87}$Rb atoms with $\mu = \frac{1}{2} \mu_B$,
respectively, where $\mu_B$ is the Bohr magneton.
Such a two-component mixture with a large difference in magnetic
moments is suitable for dipolar pattern formation.
To understand this, for simplicity, suppose a uniform system with
$g_{11} \simeq g_{22} \simeq g_{12}$.
In this case, in the absence of the DDI, the total density $|\psi_1|^2
+ |\psi_2|^2$ is almost constant ($n$), and we rewrite the contact
interaction terms for component 1 in Eq.~(\ref{GP}) as $(g_{11}
|\psi_1|^2 + g_{12} |\psi_2|^2) \psi_1 = (g_{11} - g_{12}) |\psi_1|^2
\psi_1 + g_{12} n \psi_1$, which indicates that the effective
interaction for component 1, i.e., that for quasi-spin, is $g_{11} -
g_{12}$.
Thus, the contact interaction is effectively reduced compared with the
single-component system, and hence the relative strength of the DDI is
increased.
On the other hand, the dipolar interaction part can similarly be
rewritten as $(\mu_1 |\psi_1|^2 + \mu_2 |\psi_2|^2) \psi_1 = (\mu_1 -
\mu_2) |\psi_1|^2 \psi_1 + \mu_2 n \psi_1$, and hence the effective
magnetic moment for the quasi-spin is $\mu_1 - \mu_2$.
This is the reason why we are using atomic species with $\mu_1 \gg
\mu_2$.
If $\mu_1 \simeq \mu_2$, the effective magnetic moment for the
quasi-spin vanishes, and only total-density patterns appear, as in
Ref.~\cite{scheiermann2022}.

We numerically solve the three-dimensional GP equation~(\ref{GP})
using the pseudospectral method.
The spatial discretization is typically $dx = dy = dz \simeq 0.3$
$\mu{\rm m}$ and the time step is $dt \simeq 10$ $\mu{\rm s}$.
To obtain the ground state, we solve the imaginary-time evolution.
The intracomponent scattering lengths are fixed to $a_{11} = 83
a_B$~\cite{politi2022} and $a_{22} = 100.4 a_B$~\cite{kempen2002},
where $a_B$ is the Bohr radius.
The numbers of atoms are $N_1 = 10^5$ and $N_2 = 2 \times 10^5$.

\begin{figure}[tb]
\includegraphics[width=8.5cm]{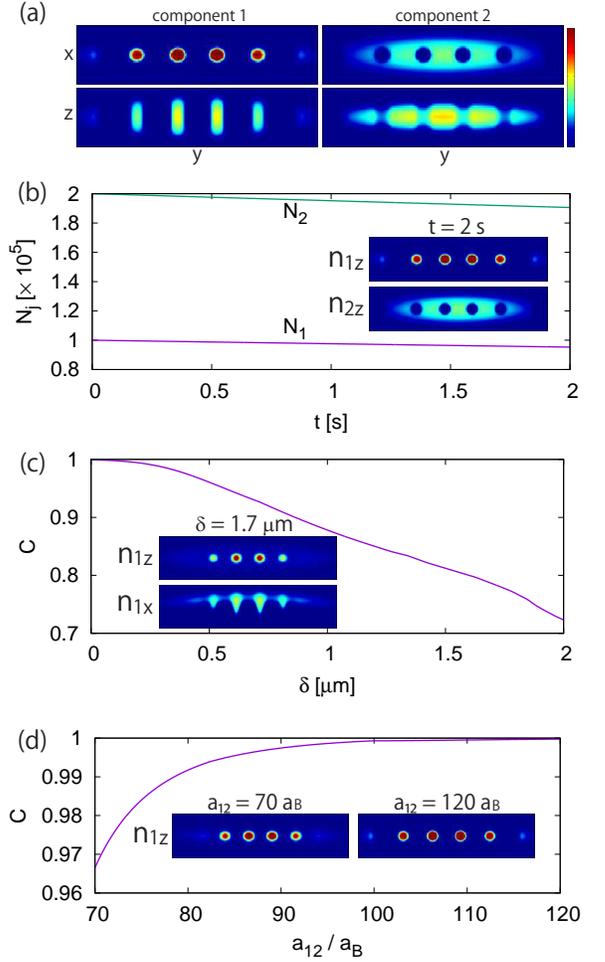}
\caption{
  (a) Integrated density distributions $n_{jz}(y, x) = \int dz
  |\psi_j(\bm{r})|^2$ and $n_{jx}(y, z) = \int dx |\psi_j(\bm{r})|^2$
  of the ground state for $N_1 = 10^5$, $N_2 = 2 \times 10^5$, $a_{12}
  = 100 a_B$, and $\delta = 0$.
  (b) Time evolution of the number $N_j$ of atoms in component $j$
  with the three-body recombination loss, where the state in (a) is
  used as the initial state.
  The insets show $n_{1z}(y, x)$ and $n_{2z}(y, x)$ at $t = 2$ s.
  (c) Dependence of the contrast $C$ on the sag $\delta$.
  The insets show $n_{1z}(y, x)$ and $n_{1x}(y, z)$ for $\delta = 1.7$
  $\mu{\rm m}$.
  (d) Dependence of the contrast $C$ on $a_{12}$.
  The insets show $n_{1z}(y, x)$ for $a_{12} = 70 a_B$ and $120 a_B$.
  In (c) and (d), other parameters are the same as those in (a).
  The images are $77 \mu{\rm m} \times 19.3 \mu{\rm m}$ in size.
  The color bar scales from 0 to $1.3 \times 10^3 \mu{\rm m}^{-2}$.
}
\label{f:1d}
\end{figure}
First, we consider a system confined in a cigar-shaped trap with
frequencies $\omega_{1x} = 2\pi \times 25$ Hz, $\omega_{1y} = 2\pi
\times 5$ Hz, $\omega_{1z} = 2\pi \times 35$ Hz, and
$\omega_{2x, y, z} = 1.8 \omega_{1x, y, z}$.
Figure~\ref{f:1d}(a) shows a typical droplet pattern formed in the
two-component BEC.
The erbium BEC splits into several droplets with a cylindrical shape,
which align in the $y$ direction, and the rubidium BEC surrounds these
droplets.
We note that the droplets are stable without the LHY correction.
In fact, the atomic density in each droplet is $\simeq 2 \times 10^{20}$
${\rm m}^{-3}$, which is smaller than that in the LHY
droplet of $^{166}$Er~\cite{chomaz2016}.
The three-body recombination loss is thus suppressed and the droplets
can survive for longer time.

We confirm the long lifetime of the droplet lattice by simulating the
real-time evolution including the atomic loss.
The three-body recombination loss can be taken into account by adding
the term $-i \hbar L_j^{(3)} |\psi_j|^4 \psi_j / 2$ to the right-hand
side of Eq.~(\ref{GP}).
The loss rate $L_1^{(3)}$ for $^{166}$Er largely depends on the
scattering length near the Feshbach resonance, and for the present
value of $a_{11} = 83 a_B$, the loss rate is given by $L_1^{(3)}
\simeq 2 \times 10^{-42}$ ${\rm m}^6/{\rm s}$~\cite{chomaz2016}.
By contrast, in the experiment of LHY droplets, the scattering length
must be smaller, where the loss rate is much larger (e.g., $L_1^{(3)}
\simeq 6 \times 10^{-41}$ ${\rm m}^6/{\rm s}$ for $a_{11} \simeq 50
a_B$~\cite{chomaz2016}).
Thus, our system has an advantage not only of the low density but also
of the small loss-rate coefficient, compared with the case of LHY
droplets.
The loss rate for $^{87}$Rb is given by $L_2^{(3)} \simeq 5.8 \times
10^{-42}$ ${\rm m}^6/{\rm s}$~\cite{burt1997}.
We neglect three-body processes in which both $^{166}$Er and
$^{87}$Rb atoms participate, since the two components are almost 
separate from each other, as seen in Fig.~\ref{f:1d} (a).
Figure~\ref{f:1d}(b) shows the time evolution of the number of atoms
$N_j$ starting from the droplet lattice state in Fig.~\ref{f:1d}(a).
The decay in the number of atoms is only a few percent for both
components, and the droplet lattice is maintained even at $t = 2$ s,
indicating a much longer lifetime compared with the LHY droplets.

The appearance of a crystal-like structure by itself is not decisive
evidence of a supersolid state.
For supersolidity, the system must allow superflow between the
droplets.
A bottleneck for the superfluid current is caused by the density
minimum along the pass between adjacent droplets.
We therefore define the minimum density as $n_{\min} = \min_y
[\max_{xz} |\psi_1(\bm{r})|^2]$, where the maximum in the $x$-$z$
plane is taken first and the minimum along $y$ between the droplets is
chosen.
The supersolidity can be quantified by the contrast $C$ defined by
\begin{equation}
C = \frac{n_{\max} - n_{\min}}{n_{\max} + n_{\min}},
\end{equation}
where $n_{\max}$ is the peak density of the droplets.
The droplet lattice exhibits no supersolidity for $C = 1$, and the
degree of supersolidity increases with decreasing $C$.
The value of $C$ for the state in Fig.~\ref{f:1d}(a) is larger than
0.99 and the degree of supersolidity is low.

To decrease the contrast $C$ and enhance the supersolidity, we
introduce a sag $\delta$ between the two components.
Figure~\ref{f:1d}(c) shows $C$ as a function of $\delta$.
We see that $C$ decreases with $\delta$, which indicates that the
supersolidity is enhanced by the sag.
This is because the droplets of component 1 are lifted upward for
$\delta > 0$, and at the region near the upper edge, the adjacent
droplets are strongly linked with each other, as seen in $n_{1x}(y,
z)$ in Fig.~\ref{f:1d}(c).
Thus, in the two-component system, the supersolidity can be controlled
by the sag $\delta$ between the two components.
When $\delta$ is larger, however, the number of droplets
decreases and they disappear.
In Fig.~\ref{f:1d}(c), the number of droplets changes to two for
$\delta \gtrsim 1.8$ $\mu{\rm m}$.
Since the precise value of the intercomponent scattering length
$a_{12}$ between $^{166}$Er and $^{87}$Rb is unknown at present, we
study the $a_{12}$ dependence, which is shown in Fig.~\ref{f:1d}(d).
The droplet pattern is stable for a wide range of $a_{12}$, and the
present results are not very sensitive to the unknown parameter
$a_{12}$.
For $a_{12} \lesssim 65 a_B$, the droplet pattern vanishes for the
condition in Fig.~\ref{f:1d}(d).

\begin{figure}[tb]
\includegraphics[width=8.5cm]{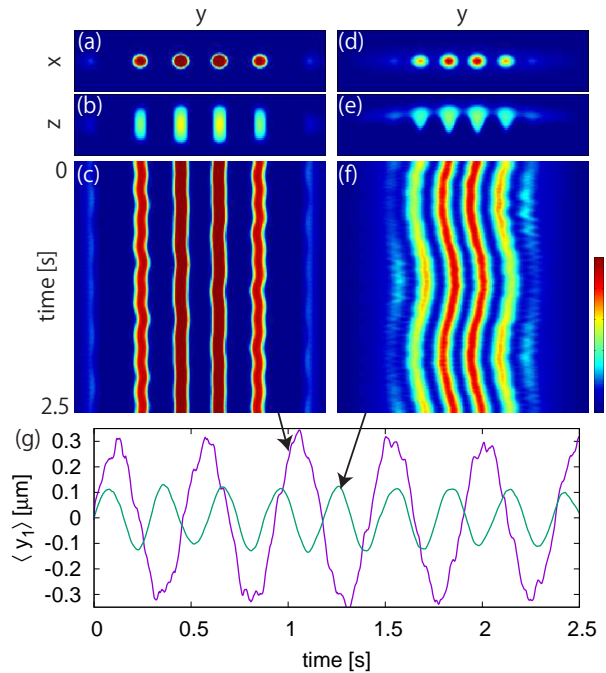}
\caption{
  Out-of-phase Goldstone mode in a one-dimensional supersolid.
  (a)-(c) Sag is absent, with $\delta = 0$ and (d)-(f) sag is present,
  with $\delta = 2.2$ $\mu{\rm m}$.
  Integrated density profiles (a), (d) $n_{1z}(y, x) = \int dz
  |\psi_1(\bm{r})|^2$ and (b), (e) $n_{1x}(y, z) = \int dx
  |\psi_1(\bm{r})|^2$ of component 1 of the ground states, where the
  images are $77 \mu{\rm m} \times 19.3 \mu{\rm m}$.
  (c), (f) Time evolution of the integrated density profile $n_1(y, t)
  = \int dx dz |\psi_1(\bm{r}, t)|^2$ of component 1, where the
  three-body loss is included.
  The initial phase is imprinted as in Eq.~(\ref{phase}) with $\phi =
  0.2 \pi$.
  The color bar scales from 0 to $1.3 \times 10^3$ $\mu{\rm m}^{-2}$
  for $n_{1z}(y, x)$ and $n_{1x}(y, z)$, and to $5.6 \times 10^3$
  $\mu{\rm m}^{-1}$ for $n_1(y, t)$.
  (g) Time evolution of the center-of-mass position of component 1 in
  the $y$ direction, $\langle y_1(t) \rangle = \int d\bm{r} y
  |\psi_1(\bm{r}, t)|^2$,
  without and with sag.
  $\omega_{2x,y,z} = 1.6 \omega_{1x,y,z}$ in (d)-(f).
  Other parameters are the same as those in Fig.~\ref{f:1d}(a).
}
\label{f:goldstone}
\end{figure}
Next, we investigate the dynamics of the droplets to examine the
supersolidity.
We imprint the initial phase to component 1 of the ground state
$\psi_1^{\rm ground}$ as
\begin{equation} \label{phase}
  \psi_1^{\rm initial}(\bm{r}) = \left\{ \begin{array}{ll}
    \psi_1^{\rm ground}(\bm{r}) e^{-i\phi} & (y < 0) \\
    \psi_1^{\rm ground}(\bm{r}) & (y \geq 0), \end{array} \right.
\end{equation}
where we take $\phi = 0.2 \pi$.
The real-time evolution including the three-body loss is performed
starting from this initial state.
Figures~\ref{f:goldstone}(a)-\ref{f:goldstone}(c) show the initial
state and time evolution for $\delta = 0$.
Since the initial phase $\phi > 0$ in Eq.~(\ref{phase}) results in a
positive phase gradient in the $y$ direction, the system acquires a
positive initial momentum, and the center-of-mass position $\langle
y_1 \rangle$ first increases, followed by oscillation, as shown in
Fig.~\ref{f:goldstone}(g).
The amplitude $\simeq 0.3$ $\mu{\rm m}$ is small, however, and the
oscillation cannot be discerned in Fig.~\ref{f:goldstone}(c).

Figures~\ref{f:goldstone}(d)-\ref{f:goldstone}(f) show the case with
$\delta = 2.2$ $\mu{\rm m}$.
As shown in Fig.~\ref{f:goldstone}(e), due to the sag, the droplets
are connected to each other by nonzero density regions, which enhances
the supersolidity.
As a consequence, the dynamics shown in Fig.~\ref{f:goldstone}(f) is
quite different from that in Fig.~\ref{f:goldstone}(c).
Although the initial momentum given by the phase imprint is positive,
the droplet pattern first moves to the negative direction (leftward).
This counterintuitive motion is compensated by the superflow in the
$+y$ direction, and in fact, $\langle y_1 \rangle$ first increases,
followed by oscillation, as shown in Fig.~\ref{f:goldstone}(g).
Its amplitude $\simeq 0.1$ $\mu{\rm m}$ is much smaller than the
apparent amplitude of the droplet lattice ($\simeq 2$ $\mu{\rm m}$) in
Fig.~\ref{f:goldstone}(f), which is due to the superflow opposite
to the lattice movement.
We can see in Fig.~\ref{f:goldstone}(f) that when the droplet lattice
moves leftward (rightward), the density of the rightmost (leftmost)
droplet increases due to the counter superflow, which is a hallmark of
the out-of-phase Goldstone mode in a supersolid~\cite{guo2019}.
We note that such an out-of-phase Goldstone mode has a long period,
which has hindered the direct observation of oscillation dynamics in
experiments due to the short lifetime of the LHY
droplets~\cite{guo2019}.
By contrast, the present two-component system has a long lifetime and
allows the observation of long-time dynamics.

\begin{figure}[tb]
\includegraphics[width=8.5cm]{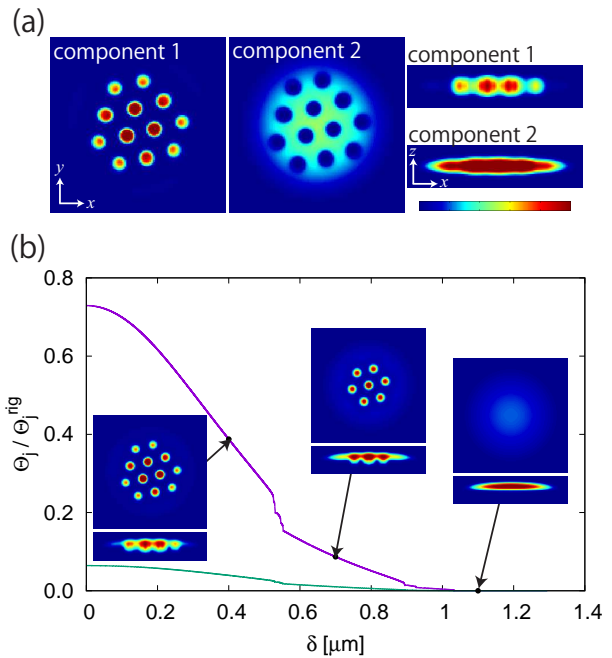}
\caption{
  Two-dimensional supersolid in an oblate trap for $a_{12} = 100 a_B$.
  (a) Integrated density profiles $n_{jz}(x, y) = \int dz
  |\psi_j(\bm{r})|^2$ and $n_{jy}(x, z) = \int dy |\psi_j(\bm{r})|^2$
  of components 1 and 2.
  (b) Normalized moment of inertia $\Theta / \Theta_{\rm rig}$ as a
  function of the sag $\delta$.
  The insets show $n_{1z}(x, y)$ and $n_{jy}(x, z)$.
  The sizes of the images are $55.2 \mu{\rm m} \times 55.2 \mu{\rm m}$
  for $n_{jz}(x, y)$ and $55.2 \mu{\rm m} \times 13.8 \mu{\rm m}$ for
  $n_{jy}(x, z)$.
  The color bar scales from 0 to $6.4 \times 10^2$ $\mu{\rm m}^{-2}$
  for $n_{jz}(x, y)$ and to $1.1 \times 10^3$ $\mu{\rm m}^{-2}$ for
  $n_{jy}(x, z)$.
}
\label{f:2d}
\end{figure}
We move on to a two-dimensional supersolid produced in an oblate system.
We consider a system confined in a trap with frequencies, $\omega_{1x}
= \omega_{1y} = 2\pi \times 9$ Hz, $\omega_{1z} = 66$ Hz, and
$\omega_{2x, y, z} = 1.5 \omega_{1x, y, z}$.
Figure~\ref{f:2d}(a) shows the ground state without sag ($\delta =
0$).
The column densities integrated in the $z$ direction clearly show a
triangular lattice of erbium droplets surrounded by the rubidium
condensate.

To study the supersolidity, we rotate the system about the $z$ axis
and calculate the moment of inertia $\Theta_j$.
We add a term $-\Omega L_z \psi_j$ to the right-hand side of
Eq.~(\ref{GP}) with a small $\Omega$ and solve the imaginary-time
evolution to obtain the ground state in the rotating frame, where
$\Omega$ is the rotation frequency of the system and $L_z = -i \hbar
(x \partial_y - y \partial_x)$ is the angular momentum operator in the
$z$ direction.
The moment of inertia for component $j$ is defined as $\Theta_j =
\lim_{\Omega \rightarrow 0} \int d\bm{r} \psi_j^*(\bm{r}) L_z
\psi_j(\bm{r}) / \Omega$.
To quantify the supersolidity, we normalize $\Theta_j$ by the moment of
inertia for rigid-body rotation, $\Theta_j^{\rm rig} = \int d\bm{r}
(x^2 + y^2) |\psi_j(\bm{r})|^2$, and thus $\Theta_j
/ \Theta_j^{\rm rig}$ can be regarded as the fraction of the classical
moment of inertia.
The quantity $\Theta_j / \Theta_j^{\rm rig}$ vanishes for an isotropic
state without droplets, and increases when droplets are
formed~\cite{roccuzzo2020, roccuzzo2022}.

We change the sag $\delta$ and calculate
$\Theta_j / \Theta_j^{\rm rig}$ for the stationary state for each
$\delta$, which is shown in Fig.~\ref{f:2d}(b).
When $\delta = 0$, the droplets are almost isolated from each other,
and they behave as a classical crystal, giving the largest $\Theta_1 /
\Theta_1^{\rm rig} \simeq 0.7$.
This is still smaller than unity due to the existence of the
superfluid halo around the droplets~\cite{baillie2018}.
The moment of inertia for component 2 is much smaller than that for
component 1, since the porous pattern, as shown by $n_{2z}(x, y)$ in
Fig.~\ref{f:2d}(a), does not much prevent superflow and the
nonclassical nature is maintained.
As the sag $\delta$ is increased, the droplet pattern is unchanged
up to $\delta \simeq 0.5$ $\mu{\rm m}$, while $\Theta_1 /
\Theta_1^{\rm rig}$ decreases to $\simeq 0.3$.
This indicates that the Josephson links between the droplets are
increased by the sag, in a manner similar to
Fig.~\ref{f:goldstone}(e), and hence the supersolidity is enhanced.
When $\delta$ is further increased, the number of droplets decreases
and eventually the isotropic state without droplets with $\Theta_1 =
0$ is reached.

In conclusion, we have proposed a two-component BEC of
a $^{166}$Er-$^{87}$Rb mixture to study supersolidity, where the LHY
correction and the Feshbach control are not needed for the formation
and stabilization of the droplet lattice state.
The lifetime of the system is thus prolonged significantly, which
enables us to observe long-time dynamics, such as an out-of-phase
Goldstone mode.
The coherence between droplets can be controlled by the difference
$\delta$ between the trap centers of the two components.
The present results may also be realized by europium
atoms~\cite{miyazawa2021} having the same magnetic moment as erbium.

This work was supported by JSPS KAKENHI Grant Number JP20K03804.


%

\end{document}